\documentclass[11pt,amsmath,a4]{article}

\usepackage{graphicx}
\usepackage{textcomp}
\usepackage{amsmath}
\usepackage{multicol}
\usepackage{authblk}
\usepackage{hyperref}
\hypersetup{
    colorlinks=true,
    linkcolor=blue,
    filecolor=magenta,      
    urlcolor=cyan,
    pdfpagemode=FullScreen,
    }
    
\usepackage{url}
\usepackage{lscape}

\title{Characterization of a continuous muon source for the Muon-Induced X-ray Emission (MIXE) Technique}

\author{Sayani Biswas$^{1}$*, Lars Gerchow$^{1}$*, Hubertus Luetkens$^{1}$, Thomas Prokscha$^{1}$, Aldo Antognini$^{1,2}$, Niklaus Berger$^{3}$, Thomas Elias Cocolios$^{4}$, Rugard Dressler$^{1}$, Paul Indelicato$^{5}$, Klaus Jungmann$^{6}$, Klaus Kirch$^{1,2}$, Andreas Knecht$^{1}$, Angela Papa$^{1,7}$, Randolf Pohl$^{8}$, Maxim Pospelov$^{9,10}$, Elisa Rapisarda$^{1}$, Peter Reiter$^{11}$, Narongrit Ritjoho$^{1,2}$, Stephanie Roccia$^{12}$, Nathal Severijns$^{4}$, Alexander Skawran$^{1,2}$, Stergiani Marina Vogiatzi$^{1,2}$, Frederik Wauters$^{3}$, Lorenz Willmann$^{6}$, and Alex Amato$^{1}$*}

\affil{$^{1}$ Paul Scherrer Institute, 5232 Villigen PSI, Switzerland}
\affil{$^{2}$ Institute for Particle Physics and Astrophysics, ETH Z\"{u}rich, 8093 Z\"{u}rich, Switzerland}
\affil{$^{3}$ PRISMA+ Cluster of Excellence and Institute of Nuclear Physics, Johannes Gutenberg Universit\"{a}t Mainz, Mainz, Germany}
\affil{$^{4}$ KU Leuven, Instituut voor Kern-en Stralingfysica, B$-$3001 Leuven, Belgium}
\affil{$^{5}$ LKB Paris, France}
\affil{$^{6}$ VSI, University of Groningen, Groningen, The Netherlands}
\affil{$^{7}$ Departimento di Fisica, Univerist\`{a} di Pisa and INFN sez. Pisa, Largo B. Pontecorvo 3, 56127 Pisa, Italy}
\affil{$^{8}$ Johannes Gutenberg Universit\"{a}t Mainz, Institute of Physics, QUANTUM and Cluster of Excellence PRISMA+, Mainz, Germany}
\affil{$^{9}$ University of Victoria, Victoria, Canada}
\affil{$^{10}$ Perimeter Institute, Waterloo, Canada}
\affil{$^{11}$ Institut f\"{u}r Kernphysik, Universit\"{a}t zu K\"{o}ln, Z\"{u}lpicher Str. 77, D$-$50937 K\"{o}ln, Germany}
\affil{$^{12}$ Universit\'{e} Grenoble Alpes, CNRS, Grenoble INP, LPSC-IN2P3, 38026 Grenoble, France}

\begin{document}
\onecolumn
\maketitle

\newpage

\abstract{The toolbox for material characterization has never been richer than today.  
Great progress with all kinds of particles and interaction methods provide access to nearly all properties of an object under study.
However, a tomographic analysis of the subsurface region remains still a challenge today.
In this regard, the Muon-Induced X-ray Emission (MIXE) technique has seen rebirth fueled by the availability of high intensity muon beams.
We report here a study conducted at the Paul Scherrer Institute (PSI).
It demonstrates that the absence of any beam time-structure leads to low pile-up events and a high signal-to-noise ratio (SNR) with less than one hour acquisition time per sample or data point.
This performance creates the perspective to open this technique to a wider audience for the routine investigation of non-destructive and depth-sensitive elemental compositions, for example in rare and precious samples.
Using a hetero-structured sample of known elements and thicknesses, we successfully detected the characteristic muonic X-rays, emitted during the capture of a negative muon by an atom, and the gamma-rays resulting from the nuclear capture of the muon, characterizing the capabilities of MIXE at PSI. 
This sample emphasizes the quality of a continuous beam, and the exceptional SNR at high rates. 
Such sensitivity will enable totally new statistically intense aspects in the field of MIXE, e.g. elemental 3D-tomography and chemical analysis.
Therefore, we are currently advancing our proof-of-concept experiments with the goal of creating a full fledged permanently operated user station to make MIXE available to the wider scientific community as well as industry.}

\section{Introduction}

Elemental analysis of materials, qualitative and quantitative, is used in a broad range of scientific fields.
Depending on the type of application, several elemental analysis techniques have been developed over the years, broadly categorized into two types: destructive and non-destructive. 
The different destructive methods include Auger Electron Spectroscopy~\cite{la06}, Scanning Electron Microscopy/Energy Dispersive X-ray Spectrometry~\cite{ne13}, Secondary Ion Mass Spectrometry~\cite{co81}, Inductive Coupled Plasma Atomic Emission Spectroscopy~\cite{ba81} and Inductive Coupled Plasma Mass Spectroscopy~\cite{da12}. 
With these destructive techniques, one can study trace elements (down to parts per quadrillion (ppq) levels) near the surface of the material, but this comes with the cost that the investigated sample cannot be retained in its original form. 
Thus, such techniques are not appropriate with rare and precious samples for which even a partial destruction is excluded. 
The different non-destructive techniques include X-ray Photoelectron Spectroscopy~\cite{az17}, X-Ray Fluorescence (XRF)~\cite{ph92}, Proton-induced X-ray Emission (PIXE)~\cite{he77}, Rutherford Backscattering Spectrometry (RBS)~\cite{he84}, Nuclear Reaction Analysis (NRA)~\cite{am84}, Prompt Gamma-ray neutron Activation Analysis (PGAA)~\cite{he09}, Neutron Activation Analysis (NAA)~\cite{dr58}, and Neutron Depth Profiling~\cite{tr18}.
All these non-destructive techniques, with the exception of PGAA and NAA, are able to provide information from near (i.e., up to $\sim$ 10 micrometer) the surface of the material only.
On the other hand, the PGAA and NAA techniques are bulk measurements and not depth-sensitive, and the sensitivity is strongly isotope dependent. 
The technique of Muon-Induced X-ray Emission (MIXE), a non-destructive technique, which was developed more than 40 years ago~\cite{ko81, re78, hu76, ta73}, has recently been used extensively with pulsed muon beams for elemental analysis~\cite{te14, ni15, te17, ni18, um18, ni19, um20, hi16, br18, cl19_1, cl19_2, ha19, ar20, gr21}. 
The advantage of this technique is that it is able to probe deep into the material, up to a few millimeters, and does not lead to a severe radiation damage of the sample. 
The aim of the present manuscript is to demonstrate the performance of the MIXE technique using a continuous muon beam.
\section{The MIXE technique}

The muon is a lepton of mass $m_{\mu}=105.66$~MeV/c$^{2}$, and thus ${\sim}207$ times heavier than the electron ($m_{e}=0.511$~MeV/c$^{2}$). 
There are two kinds of muons, namely positive muons ($\mu^{+}$, antiparticle) and negative muons ($\mu^{-}$, particle). 
For research on condensed matter and material science, spin polarized $\mu^{+}$ are routinely used in different experiments such as muon spin rotation, relaxation and resonance studies ($\mu$SR) \cite{muon_spectroscopy_2021}. 
The advantage of $\mu^{+}$ is that these stop at interstitial sites in a sample, acting as localized magnetic probes by making use of the asymmetric emission of positrons when they decay.
The $\mu^{-}$ can also be used for $\mu^{-}$SR studies~\cite{su18}. 
However, as the negative muons do not stop at an interstitial site but are rather captured by the atoms of the sample, due to the electromagnetic interaction with the nucleus, they form the so-called muonic atoms. 
Hence, the interpretation of $\mu^{-}$SR data can be quite difficult, hampering a large use of this technique. 

An alternative use of $\mu^{-}$ is to study the formation process of muonic atoms. 
When a $\mu^{-}$ is captured by an atom, the resulting muonic atom is typically created in an excited state, with the muon in an muonic orbit principal quantum number $n_{\mu}{\sim}14$~\cite{na03}. 
Subsequently, the muon relaxes in a time-scale of the order of $10^{-13}$s to the lowest $n_{\mu}=1$ muonic orbit by emitting a series of so called muonic X-rays ($\mu$-X rays)~\cite{me01}.
The energy of the $\mu$-X rays, which can be used to investigate the charge radius of the nucleus~\cite{an69, kn20, kr21}, is a fingerprint of the type of atom having captured the muon. 
Hence, the determination of the $\mu$-X ray energies provides information about the presence of the different atomic species.
Correcting the intensities of the observed $\mu$-X rays, for various processes, allows to quantify the weight percentages of the elements present in the sample under study.
The processes to take into account are the branching ratio of the $\mu$-X rays, competing Coulomb capture probabilities of the different elements (generally heavier atoms have a higher capture probability) and X-ray absorption effects inside the sample.


Due to their higher mass, muons exhibit a different stopping profile in matter than X-rays or electrons. 
Their energy loss distribution is similar to that of protons, a Bragg curve.
The mean penetration depth of muons depends on the density of the target material and the incoming muon momentum~\cite{gr01}. 
This allows to deduce the elemental composition of a material deep below the surface in the range from tens of $\mu$m to mm and profiling with step sizes down to 10~$\mu$m to 100~$\mu$m, respectively.
As an example, for a muon beam with a mean momentum of 30 MeV/c (corresponding to a kinetic energy of 4.18 MeV) and a momentum distribution characterized by a momentum bite $\Delta p/p = 2\%$ ($\sigma = \Delta p$), the stopping depth 
is $260 \pm 20$~$\mu$m in a copper target. 

Also owing to the heavier mass, a $\mu$-X ray has an energy ${\sim}207$ times higher than that associated with the corresponding electronic transition (neglecting finite size effects). 
This much higher $\mu$-X ray energy and the correspondingly smaller
mass attenuation coefficient
has the consequence that the created $\mu$-X rays have a much higher probability to pass through the target material from much deeper regions than the characteristic X-rays of electronic transitions. 

In analogy to the Moseley law~\cite{mo13} for the electrons, and by assuming a point-charge nucleus, one can express the Bohr model energy of the $\mu$-X ray created between transitions between states with principal quantum numbers $n_i$ (initial) and $n_f$ (final) as:
\begin{equation}
E_{i\to f,\mu} = \frac{\bar{m_{\mu}}}{m_{e}}R_{y}(Z-S_{scr,\mu})^{2}(\frac{1}{n_{f}^{2}}-\frac{1}{n_{i}^{2}}) \simeq \frac{\bar{m_{\mu}}}{m_{e}}E_{i\to f,e} \simeq 207 \times E_{i\to f,e}~,
\label{eqn:eq1}
\end{equation}
where $\bar{m}_{\mu}$ is the reduced mass of the system formed by the muon and the nucleus and $R_{y}$ is the Rydberg constant. 
Note that by analogy with the Moseley law, we replace the atomic number $Z$ by $Z_{\text{eff}}=(Z-S_{\text{scr,}\mu})$.
We have $S_{\text{scr,}\mu}=0$ since there is no screening of the charge of the nucleus seen by the muon involved in the transition, as solely one $\mu^{-}$ is present at a given time in the sample. 
Though the $S_{\text{scr,}\mu}=0$, the electron screening also affects the energy of the $\mu$-X rays. 
While this effect can lead to energy shifts ranging from a fraction of eV to few hundreds of eV (depending on the muonic transition and the element), these are of the order of permille or less~\cite{bo82}.
The situation is however different for the case of electrons where for K$_\alpha$ transitions ($2p \rightarrow 1s$), for example, one has $S_{\text{scr,}e}=1$ as already one electron is present in the $1s$ level. 
This difference explains the first approximation in Eq.~\ref{eqn:eq1}. 
Note that the Bohr model ignores the fine structure of the levels, which is determined by relativistic effects and the spin-orbit coupling. 
Hence, all states with the same principal quantum number $n$ and different azimuthal quantum number $\ell$ have the same binding energy in this model. 
Note also that for heavy nuclei, the $\mu$-X ray energy will be smaller than the one predicted by Eq.~\ref{eqn:eq1}, as the size of the nucleus cannot be neglected anymore with respect to the characteristic radius of the muonic state. 
This deviation will be smaller for large values of $n_f$, corresponding to a large radius of the muonic state.

Concluding, by measuring the $\mu$-X ray energy one can identify the element (similar to the XRF technique), which had captured the muon. 
However, the fundamental differences compared to the XRF technique are, that the muon momentum can be tuned to implant the muons at controlled depths deep inside the material (from tens of micrometers to several millimeters), and that the created $\mu$-X rays have enough energy to escape the material and being detected.\footnote{We note here that for pure iron an X-ray created by the muonic transition K$_{\alpha 2}$ will have a range three orders of magnitude larger than the corresponding X-ray created by an electronic transition.}    
This technique is referred to as the Muon-Induced X-ray Emission (MIXE) technique. 
MIXE can be applied to carry out non-destructive elemental analysis deep inside a material, which is not possible by XRF or PIXE.

Once the muon has reached its ground state, i.e. $n_f=1$, it either decays or gets captured by the nucleus. 
The respective branching probabilities will depend on the overlap between the muonic and nuclear wavefunctions. 
The probability of the 
capture by the nucleus increases with $Z$ through the reaction
\begin{equation}
\mu^{-} + p \longrightarrow n + \nu_{\mu}~.
\label{eqn:eq2}
\end{equation}
This leads to a decrease of the muon lifetime, as the muon lifetime $\tau$ is given by
\begin{equation}
\frac{1}{\tau} = \frac{1}{\tau_\mathrm{decay}} + \frac{1}{\tau_\mathrm{nuclear~capture}}~.
\label{eqn:eq3}
\end{equation}
For low $Z$ muonic atoms the muon lifetime is essentially identical to that of the free muon ($\sim 2.2$~$\mu$s), whereas for heavy atoms it can be lower than 100~ns~\cite{su87}.

The capture of the $\mu^{-}$ by the nucleus (Eq.~\ref{eqn:eq2}) results in the formation of a $Z-1$ nucleus in an excited state, which will relax through different possible processes, for which some are connected with the emission of characteristic gamma-rays. 
By determining the energy of the gamma-rays, one obtains an additional confirmation of the type of atom having captured the $\mu^{-}$. 
We note here that the possible nuclear capture of the muon (i.e. a muon not experiencing a normal decay) may lead to the formation of a radioactive nucleus and therefore questioning the claim of MIXE as a non-destructive method. 
We nevertheless note that a normal experiment requires about $10^6$ muons implanted into the sample, and depending on the nucleus, a fraction of these will be captured by it.
Such a low number of possibly created radioactive nuclei is negligible in comparison to the  number of atoms present in a typical sample, which is of the order of the Avogadro number. 

Presently, there are multiple applied research muon beam facilities in the world in operation: (i) Paul Scherrer Institute (PSI), Switzerland; (ii) ISIS, Rutherford Appleton Laboratory (RAL), United Kingdom; (iii) TRI Univeristy Meson Facility (TRIUMF), Canada; (iv) Japan Proton Accelerator Research Complex, MUon Science Establishment (J-PARC MUSE) and (v) MUon Science Innovative Channel (MuSIC), Japan.
The interested reader is referred to comprehensive review of Hillier et al.~\cite{hi22} about accomplishments, present research and future capabilities at the institutes listed above.
The cyclotron-based high beam duty factor facilities at PSI, TRIUMF, and MuSIC provide continuous muon beams to the experimental areas. 
The continuous muon beams allow detecting each incident muon impinging on the experimental target one-at-a-time.
%
%
In our experiments, the next muon arrives on average after $>~10$ muon lifetimes, allowing sufficient time for the previous muon to decay or be captured by the nucleus and therefore minimizing event pile-up in the germanium detectors, which are used to detect the $\mu$-X rays.

The synchrotron-based J-PARC MUSE and ISIS RAL facilities, with lower beam duty factor produce pulsed muon beams at 25~Hz and 50~Hz, respectively. 
The instantaneous muon rates during these short pulses (having a characteristic width of the order of 100~ns) are so high that the sample is hit by thousands of muons 
in this short time window, generating a lot of pile-up events in the Ge detectors (which have response times of the order of few microseconds). 
Although pile-up events can be identified and rejected, the consequent loss of events drastically limits the detection of relatively weak features in a high background of unwanted events. 
In the following, we will demonstrate that a continuous muon source, such as PSI, is the ideal place to perform fast and precise MIXE measurements and that within a few minutes of data collection one obtains a clear elemental identification. 
This enables systematic measurements of a large number of samples in a short period of time.
We would like to stress that the use of the MIXE technique to perform elemental analysis at PSI already started decades ago \cite{ha76,da78,eg81}, but was hampered by the available muon rates at that time.
Nevertheless, already very promising results were obtained.
Nowadays, the beam intensity is at least a factor of 20 higher, opening new application possibilities of this technique.
\section{Experimental Details and Results}

\begin{figure}[]
\includegraphics[angle=0,width=1.0\linewidth]{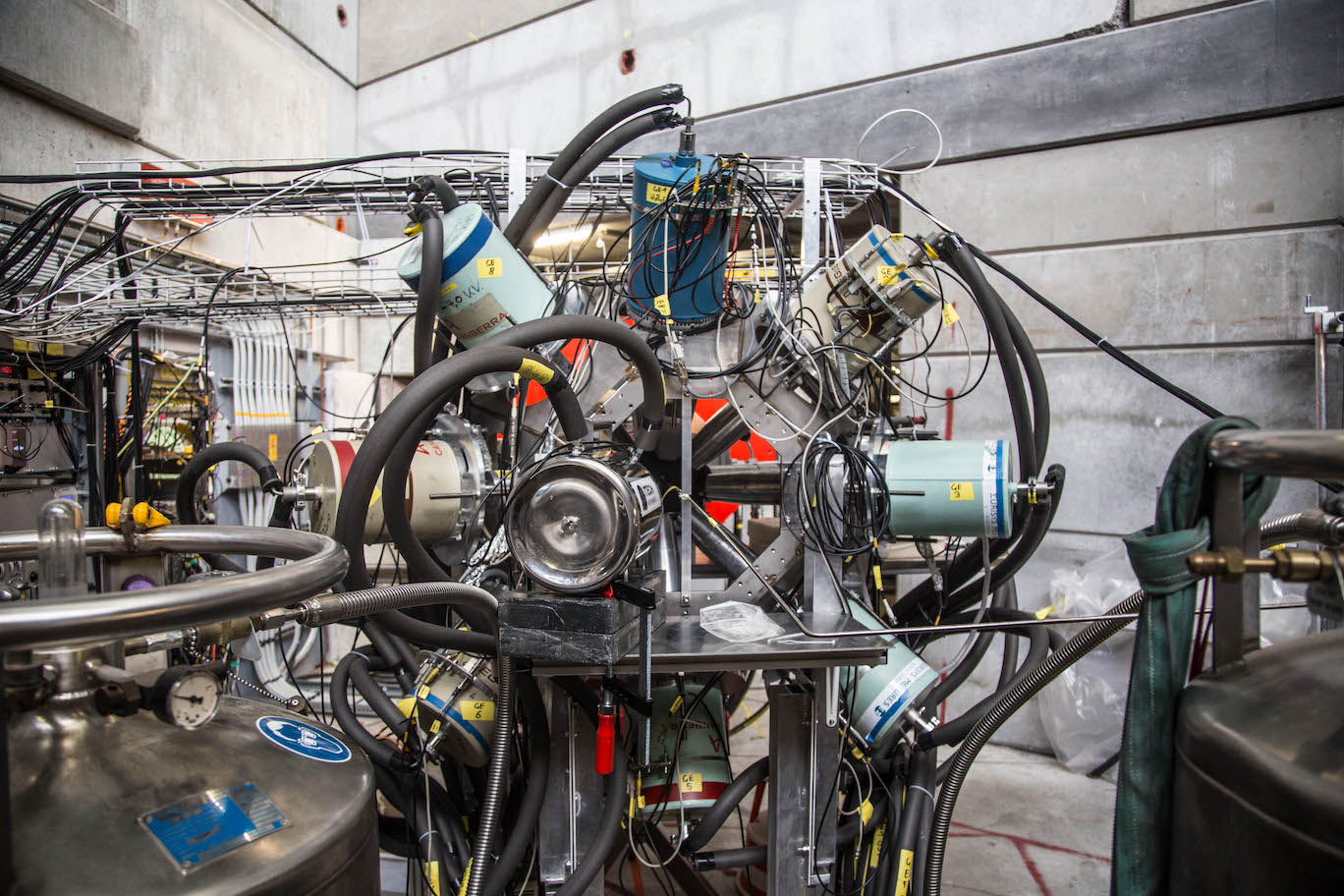}
\caption{\label{fig:fig0} A photograph of the experimental setup showing the assembly of the detectors.}
\end{figure}

\begin{figure}[]
\includegraphics[angle=0,width=1.0\linewidth]{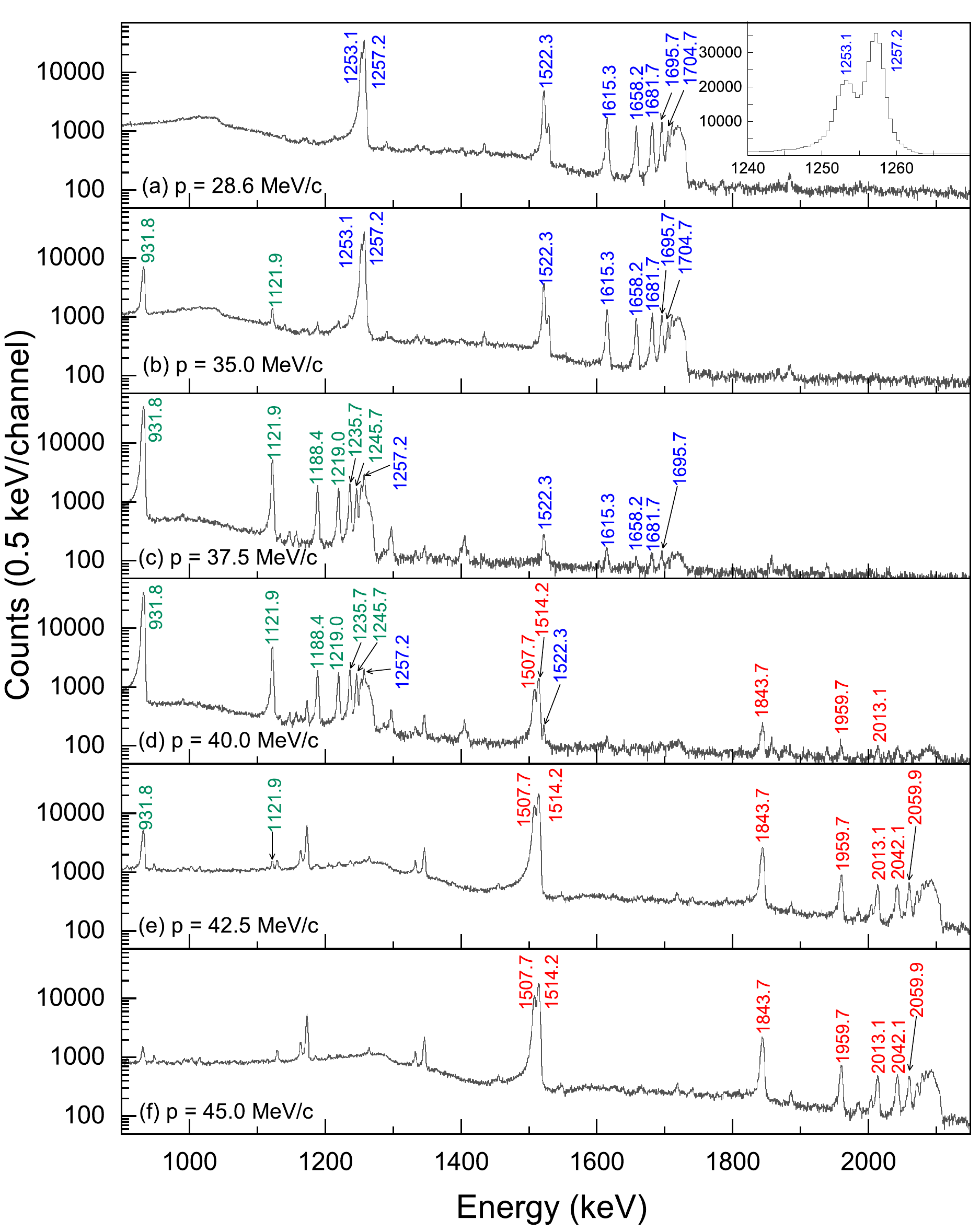}
\caption{\label{fig:fig1} The muonic X-ray spectrum in the energy range 900 to 2150 keV, from all the twelve coaxial Ge detectors, for the six different muon momenta. The data accumulation time for each momentum was $\sim$ 30 min. The $\mu$-X rays of Fe, Ti and Cu are labelled in blue, green, and red colors, respectively. The energies of the $\mu$-X rays were taken from~\cite{muxrays_jinr,me06}. The inset in panel (a) shows the zoomed portion of the spectrum (from 1240 to 1270~keV), to emphasize the fine structure of the muonic $K_{\alpha}$ lines in Fe.}
\end{figure}

The MIXE experiment was performed using the $\pi$E1 beamline, fed by the High Intensity Proton Accelerator (HIPA) \cite{Grillenberger:2021HIPA} of the Paul Scherrer Institute (PSI), Switzerland. 
This beamline is capable of delivering positive or negative muons~\cite{piE1_website}. 
With this beamline, we could achieve typical muon rates of $\sim$1.5 kHz and $\sim$60 kHz at momentum $p$~=~20 and 45~MeV/c, respectively, with momentum bite $\Delta p/p$~=~2\% ($\sigma = \Delta p$). 
We also observed a pile-up of $\sim 1\%$ with the muon rates of $\sim$60 kHz at $p$~=~45 MeV/c and momentum bite of 2\%.
Even at momentum bite of 0.5\%, we could obtain reasonable muon rates, which would be ideal for the investigation of thin samples.
As the $\pi$E1 beamline of the HIPA accelerator complex at PSI delivers a continuous beam, one has to detect the individual arrival time of each muon using a muon entrance detector, made of polyvinyltoluene (BC-400), and of ${\sim}200~{\mu}$m thickness. 
In addition, a muon veto counter with a 18~mm (diameter) hole, made of polyvinyltoluene (BC-400) is placed just before the muon entrance counter.
The experimental apparatus (see Fig.~\ref{fig:fig0} and ~\cite{alex_thesis}) was originally developed for the muX experiment~\cite{muX}. It consisted of a planar detector and a stand-alone coaxial detector (efficiency, $\epsilon\sim$70\%) from PSI; a stand-alone coaxial detector ($\epsilon\sim$75\%) from KU Leuven; one MINIBALL cluster module, with three detector crystals ($\epsilon\sim$60\% each), from KU Leuven~\cite{wa13}; seven compact coaxial detectors ($\epsilon\sim$60\%) from the IN2P3/STFC French/UK Ge Pool~\cite{pool_website}; and a low-energy detector from ETH Z\"{u}rich.
So, in total we had twelve coaxial and two low-energy Ge detectors. 
Out of the above-mentioned detectors, two of these had a BGO anti-Compton shield.
The energy and efficiency calibrations for this detector setup was done using the standard radioactive sources of $^{152}$Eu, $^{88}$Y, $^{60}$Co and $^{137}$Cs.
The target consisted of a three-layered sandwich sample of iron (Fe), titanium (Ti), and copper (Cu) plates and was placed in vacuum. 
Each layer was $500~{\mu}$m thick with a diameter of 26.5~mm. 
There were two main purposes of measuring such a sandwich sample consisting of pure elements: (i) to demonstrate the possibility of measuring the different $\mu$-X rays and gamma-rays for different elements, (ii) to determine if by changing the muon momentum, we are able to probe the element present at different depths inside the sample and how such measurements correlate with simulations for penetration depths of negative muons. 
Hence, the measurement of such a simple sample acts as proof of principle for later precise measurements on actual samples like archaeological artefacts, battery and meteorite samples where the depth-dependent elemental compositions 
are of interest.
The simple geometry of the sample was also chosen to 
benchmark our results with the ones obtained in a similar sample at the pulsed muon source ISIS~\cite{hi16}.

The sample was measured at six different muon momenta, $p$ = 28.6, 35.0, 37.5, 40.0, 42.5, and 45.0 MeV/c, with momentum bite $\Delta p/p$~=~2\% ($\sigma = \Delta p$). 
A data collection time of ${\sim}30$~min was used  for each muon momentum. 
Figure~\ref{fig:fig1} shows the $\mu$-X ray spectra of the muonic K-series, measured with all the twelve coaxial Ge detectors, for the six different muon momenta, in the energy range $900< E< 2150$~keV. 
At the lowest momentum ($p$ = 28.6 MeV/c) [Fig.~\ref{fig:fig1}(a)], all the muons stop in the first sample layer and one is able to observe the entire K-series of the $\mu$-X rays from Fe. 
In addition, the $2p_{3/2}-1s_{1/2}$ (K$_{\alpha 1}$) and the $2p_{1/2}-1s_{1/2}$ (K$_{\alpha 2}$) transitions at 1257.2~keV and 1253.1~keV are resolved, as shown in the inset of Fig.~\ref{fig:fig1}(a).

Upon increasing the muon momentum, one detects first the $\mu$-X rays from both the Fe and Ti layers [$p$ = 35.0 and 37.5~MeV/c, see Figs.~\ref{fig:fig1}(b),(c)]; then the ones from all the three Fe, Ti and Cu layers [$p$ = 40.0, see Fig.~\ref{fig:fig1}(d)].
At $p$ = 42.5 MeV/c (Fig.~\ref{fig:fig1}(e)), one looses the Fe-signal and only Ti and Cu $\mu$-X rays are observed. 
At $p$ = 45.0 MeV/c (Fig.~\ref{fig:fig1}(f)), only the Cu muonic X-rays are observed. 
It should be noted that the photopeak signal-to-noise ratio (SNR) is $\sim 20$ for the $K_{\alpha}$ line of Cu (see Fig.~\ref{fig:fig1}(f)).

To perform the analysis and estimate where the muons are stopping, we have chosen the characteristic $\mu$-X ray peaks of the three layers as follows: (i) K$_{\alpha 1}$ (1253.1 keV) and K$_{\alpha 2}$ (1257.2 keV) lines of Fe, (ii) K$_{\alpha}$ (931.8 keV) of Ti, and (iii) K$_{\alpha 1}$ (1507.7 keV) and K$_{\alpha 2}$ (1514.2 keV) lines of Cu. 
The intensity of these peaks was obtained by dividing the area under these peaks (after a proper background subtraction) by the efficiency at the corresponding energies.

Since at $p$ = 28.6 MeV/c, we see only the Fe lines, the total intensity of the K$_{\alpha 1}$ (1253.1 keV) and K$_{\alpha 2}$ (1257.2 keV) peaks is normalized to 100.
At $p$ = 35.0 MeV/c, as both Fe and Ti lines are present, the intensities of the peaks of Fe and Ti are normalized in such a manner that both sum up to 100. 
The same procedure is followed for the rest of the muon momenta.
By analyzing the relative intensities of the different peaks, we are thus able to estimate experimentally the fraction of the muon ensemble stopping in the different layers at a given muon momentum.

\begin{figure}[h!]
\includegraphics[width=1.0\columnwidth]{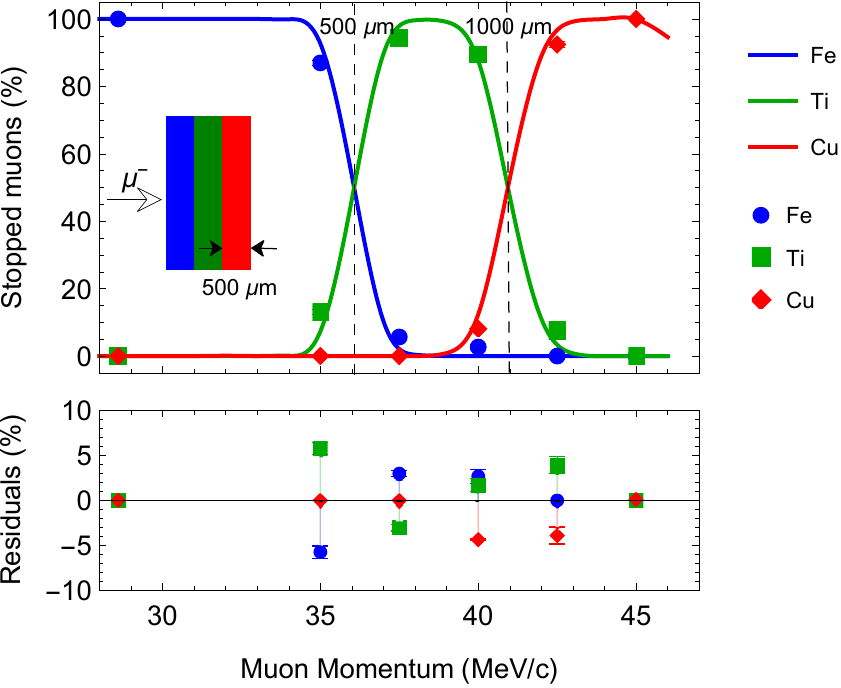}
\caption{\label{fig:fig2} (top panel) Percentage of muons stopping in each layer of the sandwich sample obtained from simulations (shown by solid lines) and experiment (points) for the three different layers Fe (blue line and blue filled circles), Ti (green line green filled squares) and Cu (red line and red filled diamonds). The inset shows a sketch of the orientation of the sandwich sample with respect to the negative muon beam and the thickness of the individual layers. Two black dashed lines are drawn at the intersection of the simulated curves. (bottom panel) The residuals (experimental-simulated) in percentage for the different data points.}
\end{figure}

To validate our approach, we compared our results with stopping profile simulations obtained with the Particle and Heavy Ion Transport code System (PHITS) simulation tool (ver. 2.88)~\cite{phits1,phits2}.
The PHITS simulations were used to represent the interaction of the negative muon beam, with a given muon momentum ($p$) and a Gaussian momentum distribution with standard deviation $\sigma = \Delta p$ (relative width $\Delta p/p = 2\%$), on the three-layered sandwich sample (Fe,Ti, and Cu) placed in vacuum, including the 200-${\mu}$m-thick muon entrance detector. 
The simulations were carried out from $p$ = 28 to 46 MeV/c, at intervals of 1 MeV/c to obtain the percentage of muons stopping in a given layer and the penetration depths of the muons. 
The interpolated simulated data points (using Hermite interpolation of order three) are shown as solid lines (blue for Fe, green for Ti, and red for Cu) in Fig.~\ref{fig:fig2}(a).
The efficiency corrected and normalized intensities (as described in the previous paragraph) of the muonic$-K_{\alpha}$ lines of Fe, 
Ti, 
and Cu 
were used to estimate the percentage of muons stopping in each layer. They are shown in the figure as blue filled circles, green filled squares and red filled diamonds, respectively. 
%
%
The difference between the experimental and the simulated points, called the residuals, at the six different muon momenta is shown in Fig.~\ref{fig:fig2}(b). From this plot, one can see that we obtain a very good agreement, within $\sim 10\%$.

\begin{figure}[]
\includegraphics[angle=0,width=1.0\linewidth]{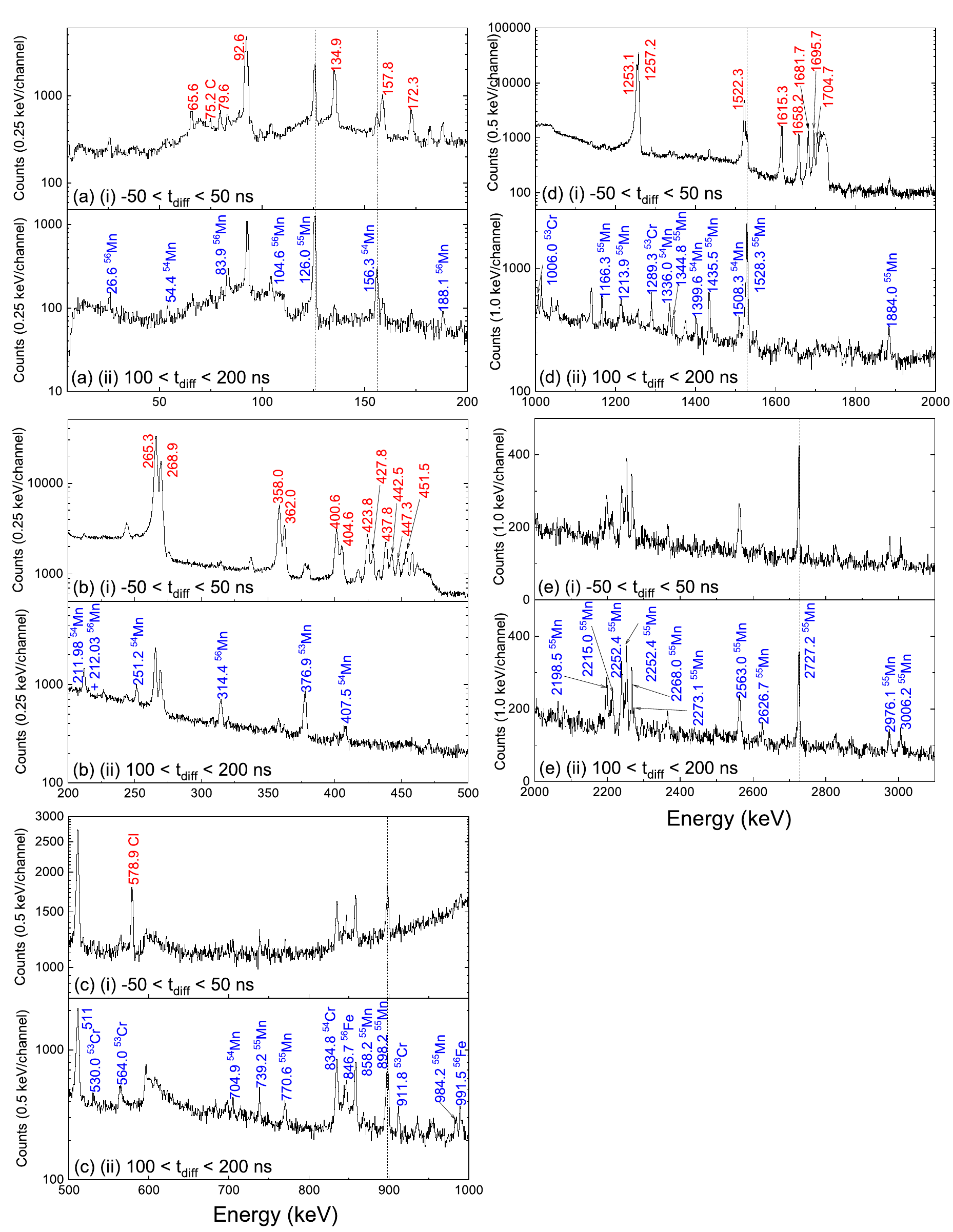}
\caption{\label{fig:fig4} The prompt muonic X-ray spectrum ((i) -50$<t_{\text {diff}}<$50 ns) and delayed gamma-ray spectrum from the nuclear capture of muons ((ii) 100$<t_{\text {diff}}<$200 ns) in the Fe layer of the three-layered sandwich sample at $p =28.6$~MeV/c within the energy ranges of (a) 0~$< E <$ 200~keV, (b) 200 $< E <$ 500~keV, (c) 500 $< E <$ 1000~keV, (d) 1000 $< E <$ 2000~keV, and (e) 2000 $< E <$ 3100~keV. The $\mu$-X ray energies are written in red, while the gamma-ray energies are in blue.}
\end{figure}


In addition to the $\mu$-X rays, 
the gamma-rays produced after the nuclear capture of muons have also been observed in the present experiment. 
We show this for the data measured at the
muon momentum of $p = 28.6$~MeV/c, 
where all the muons stop in the very first Fe layer. 
The gamma-rays, resulting from the nuclear capture in Fe, were previously studied in Ref.~\cite{ev73} and Ref.~\cite{me06}. 
In Ref.~\cite{ev73}, only six gamma-rays were detected in the Fe spectrum, resulting from the isotopes $^{56}$Fe~$({\mu^{-}},{\nu}{\gamma})$~$^{56}$Mn, $^{56}$Fe~$({\mu^{-}},{\nu}n{\gamma})$~$^{55}$Mn, $^{56}$Fe~$({\mu^{-}},{\nu}3n{\gamma})$~$^{53}$Mn, and $^{56}$Fe~$({\mu^{-}},{\nu}pn{\gamma})$~$^{54}$Cr. 
A gamma-ray at 847~keV was observed, due to the decay of the first excited state of $^{56}$Fe, in both the prompt $\mu$-X ray and the delayed gamma-ray spectra.
The presence of this 847~keV in the delayed spectrum was explained to be arising due to the inelastic scattering of the $^{56}$Fe nucleus by neutrons following the muon capture. 
In the other reference~\cite{me06}, up to forty-nine gamma-rays originating from the de-excitation of the $^{56}$Fe~$({\mu^{-}},{\nu}{\gamma})$~$^{56}$Mn, $^{56}$Fe~$({\mu^{-}},{\nu}n{\gamma})$~$^{55}$Mn, $^{56}$Fe~$({\mu^{-}},{\nu}2n{\gamma})$~$^{54}$Mn, $^{56}$Fe~$({\mu^{-}},{\nu}3n{\gamma})$~$^{53}$Mn and $^{56}$Fe~$({\mu^{-}},{\nu}pn{\gamma})$~$^{54}$Cr isotopes were observed. 
From our analysis, we observed all the gamma-rays mentioned in the previous two references.
In addition, we also observe the gamma rays from $^{56}$Fe~$({\mu^{-}},{\nu}p2n{\gamma})$~$^{53}$Cr or $^{54}$Fe~$({\mu^{-}},{\nu}p{\gamma})$~$^{53}$Cr. 

Figure~\ref{fig:fig4}(a) shows the $\mu$-X ray and gamma-ray spectra from Fe in the energy range $0 - 200$~keV, measured using the two low-energy germanium detectors. 
In order to differentiate between these two spectra, a time-gate cut has been applied. 
As the muon cascade down to the $1s$ ground state is almost instantaneous after its capture by an atom, the associated $\mu$-X ray emission will happen basically in coincidence with the muon entrance time. 
Therefore, by applying a time-cut of $-50~ns<t_{diff}<50~ns$ one
selects primarily the $\mu$-X rays of Fe (see Fig.~\ref{fig:fig4}(a)(i), where the $\mu$-X rays of the N-series at $65.6$, and $79.6$ keV and of the M-series at $92.6, 134.9, 157.8$, and $172.3$ keV are observed, already reported in Refs.~\cite{zi18, muxrays_jinr}). 
Here, $t_{diff}$ is the time difference between the muon entrance counter and the germanium detector.
Similarly, by using a time cut of 100~ns$<t_{diff}<$200~ns, i.e.~a time range where the $\mu$-X rays have already been emitted, the $\mu$-X rays are no more present in the histogram and solely the gamma-rays are obtained  (see Fig.~\ref{fig:fig4}(a)(ii)). 
Figure~\ref{fig:fig4}(a)(ii), exhibits the gamma-rays from the $^{54-56}$Mn nuclei. 
None of those have been reported in previous works, probably because these transitions are at low energy. 
The energy values of these gamma-rays were taken from the adopted level schemes in National Nuclear Data Center (NNDC)~\cite{nndc}. 
Some uncertainty was related to the existence of the 126 keV state \cite{ev73}, as the observation of a 858 keV gamma-ray was reported, which is a result of the decay from 984 keV state to the 126 keV state. 
Our present data clearly confirms the existence of the 126 keV state, since we observed the 126 keV gamma-ray.
Since the time difference between the emission of the $\mu$-X rays and the gamma-rays is very small, one would still observe the gamma-rays (but with much lower intensity compared to the $\mu$-X rays) also in the prompt $\mu$-X ray spectrum.
We observe this in Fig.~\ref{fig:fig4}(a) and some of these peaks are marked with a black dashed line.

\begin{figure}[]
\includegraphics[angle=0,width=0.8\linewidth]{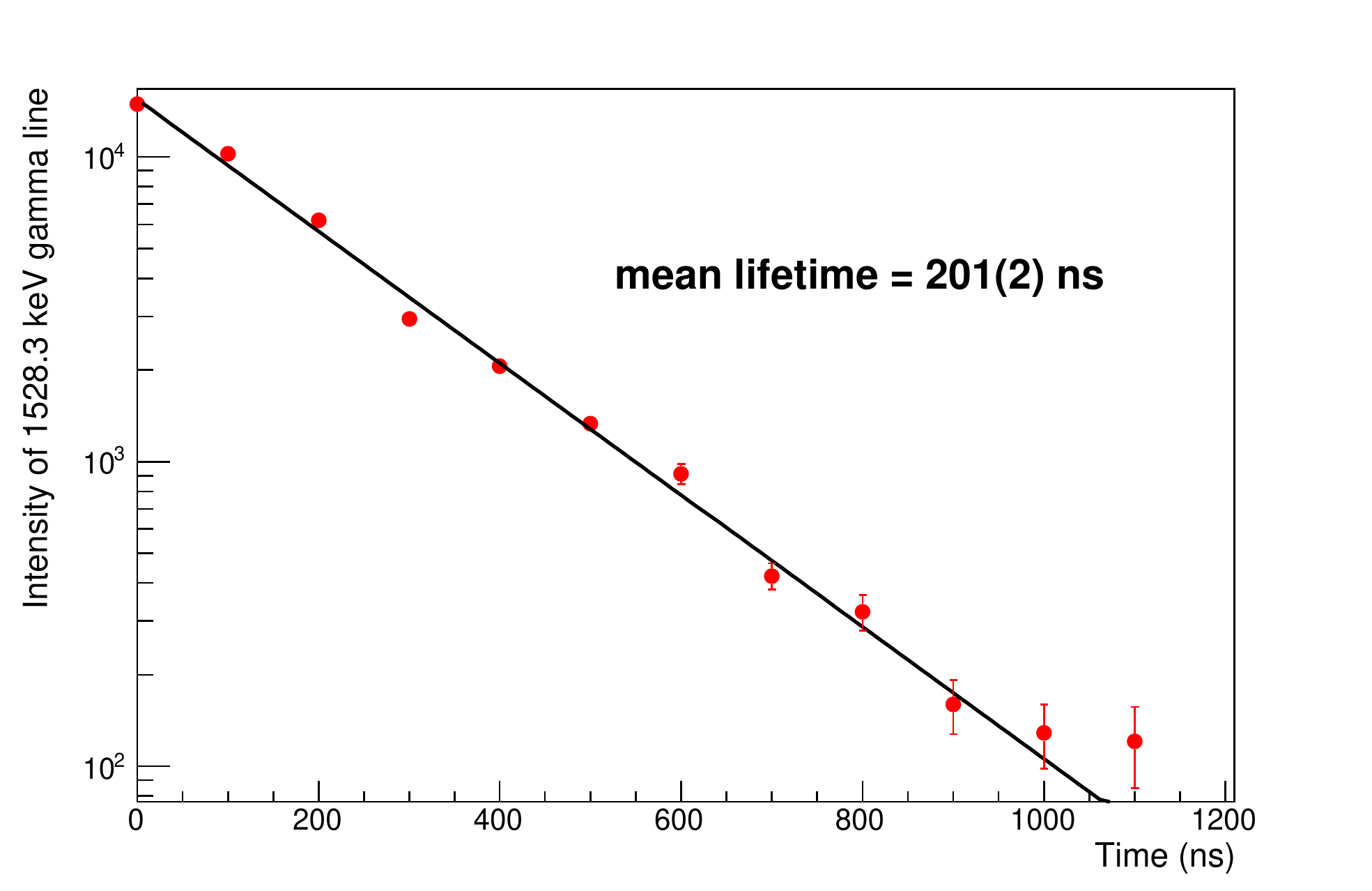}
\caption{\label{fig:fig5}Time evolution (in 100 ns wide windows) of the intensity of 1528.3 keV gamma-ray originating from the de-excitation after muon capture in the Fe nucleus, shown by red filled circles. The formula $I(t)=I[\exp(-t/\tau)]$ is fit to the data, represented by the black solid line (see also Eq.~\ref{eqn:eq3}).}
\end{figure}

Figure~\ref{fig:fig4}(b) (panels (i) and (ii)) shows the $\mu$-X ray and gamma-ray spectra from Fe in the energy range $200-500$~keV, measured using all the twelve coaxial detectors. 
We observe $\mu$-X rays of the L-series at 265.3, 268.9, 358.0, 362.0, 400.6, 404.6, 423.8, 427.8, 437.8, 442.5, 447.3, and 451.5 keV, as observed in Ref.~\cite{ha76}. 
The gamma-rays are from the $^{53,54,56}$Mn isotopes. 
Figure~\ref{fig:fig4}(c) shows the $\mu$-X ray and gamma-ray spectra from Fe in the energy range $500-1000$~keV, measured using again all the twelve coaxial detectors.
Figure~\ref{fig:fig4}(c)(i) shows a $\mu$ X-ray of $578.9$ keV, but this does not belong to Fe. This is the muonic  $K_{\alpha}$ line of Cl, which is present in the plastic (PVC), a cylindrical shaped plastic sample holder.
It is to be noted that we also observed the muonic $K_{\alpha}$ line of C at 75.2 keV (Fig.~\ref{fig:fig4}(a)(i)). 
The gamma-rays are from $^{54,55}$Mn, $^{53,54}$Cr, and $^{56}$Fe isotopes, as shown in Figure~\ref{fig:fig4}(c)(ii). 
All these gamma-rays, except those of $^{53}$Cr, were observed in the previous references.
Figure~\ref{fig:fig4}(d) exhibits the $\mu$ X-ray and gamma-ray spectra from Fe in the energy range $1000-2000$~keV. 
The $\mu$ X-rays of the K-series at $1253.06, 1257.19, 1522.3, 1615.3, 1658.2, 1681.7, 1695.7$, and $1704.7$ keV are observed (see also Ref.~\cite{me06}). 
The gamma-rays originate from $^{53, 54,55}$Mn and $^{53}$Cr isotopes. All these gamma-rays, except those of $^{53}$Cr, were observed in the previous references.

With respect to the muon lifetime ($\tau_\mu=2.2\mu$s), the emission of muonic X-rays ($\tau_\text{X-ray}<10^{-13}$~s) can be seen as in coincidence with the formation of the muonic atom, while the emission of gamma-rays ($\tau_\gamma<10^{-12}$s) in coincidence with the nuclear capture of the muon.
Therefore, the measurement of the gamma-rays as a function of time can be used to determine the muon lifetime in a particular element in the sample. 
As an example, we report here in Fig.~\ref{fig:fig5} the intensity of the 1528.3 keV gamma-ray (from muon capture in Fe at p = 28.6 MeV/c) as a function of the time of the Ge detector signals (in 100 ns wide windows). 
The fit of this plot results in a mean lifetime of muon in Fe to be 201(2) ns, which is in good agreement with the values of 201(4) \cite{se59}, 207(3) \cite{bl62}, 206.7(2.4) \cite{ec66} and 206.0(1.0)~ns \cite{su87}. 
We note here that this muon lifetime determination was not our primary goal and that a measurement as short as 30 minutes already provides an excellent precision. 
Hence, it appears that the identification of the gamma-rays and the determination of the mean lifetime of the muon act as additional proofs for the elemental identification. 
And we stress again that the determination of the mean lifetime of all elements is only possible due to the continuous muon beam, where each muon arrival time is measured. 

\section{Summary and Conclusions}

By measuring a three-layered sandwich sample of known elements, we demonstrated the advantages of using the Muon Induced X-ray Emission (MIXE) technique at continuous muon beams, available at PSI.
The presence of the characteristic muonic X-rays ($\mu$-X rays) from these known elements provides a detailed picture of the elemental composition of the different elements as a function of depth. 
The good agreement between the experimental and simulated data validates our data acquisition and analysis procedures. 
Moreover, it will allow us to utilize the simulation to design future experiments and implement a more sophisticated iterative simulation procedure, which would take into account the absorption of $\mu$-X rays in irregular shaped samples of non-uniform densities.

The possibility to detect gamma rays of the excited daughter isotopes after nuclear capture of the muon, in addition to the $\mu$-X rays, provides a further confirmation of the identification of the elements.
Relying on this success, it is clear that this technique, applied at continuous muon beams, appears as a very powerful tool to analyze different types of samples, which could be precious (as archaeological artefacts) or rare (as meteorite or "returned" samples). 
By exploiting the depth-dependent elemental identification and the short measurement times, we foresee to even perform experiments in operando devices, such as batteries, to track elemental composition changes.

\bibliography{main}
\bibliographystyle{h-physrev}

\end{document}